# Associated Functions on SO(p,q) Groups

## Bala Ali RAJABOV

*National Academy of Sciences,*
*Institute of Physics, 33 Javid avenue,*
*Baku 370143-AZERBAIJAN*

**Abstract**

Explicit expressions for associated spherical functions of SO(p,q) matrix groups are obtained using a generalized hypergeometric series of two variables.

### 1. Introduction

The SO(p,q) groups of pseudo-orthogonal matrices and their representations are broadly used in different fields of physics, particularly in quantum field theory, high-energy physics, cosmology and solid-state physics [1-2].

Matrix elements of irreducible unitary representations (IUR) of these groups, particularly spherical functions, play an essential role in the theory of representations of SO(p,q) groups. These functions in the case of SO(p, 1) groups are investigated in details [3-4]. Unlike them, the construction of spherical functions of SO(p,q), $p \geq q \geq 2$ groups has not yet been completed. N. Ya. Vilenkin's and A.P. Pavluk's publications [2], in which the "relationship between the spherical functions of matrix groups and Herz's functions of matrices argument was established [5], have been receiving remarkable attention.

The purpose of this paper is to find the spherical functions of SO(p,q) groups when $p \geq q \geq 2$. The main result consists of the fact that associated spherical functions of the groups SO(p,q), $p \geq q \geq 2$, are expressed as generalized hypergeometric Horn's functions of two variables by a unique formula, in spite of essentially different forms of the integral representations of spherical functions at $p \geq q \geq 3$ and $p \geq 2$, $q = 2$. The preliminary statements about the obtained results were made in a Report of the Academy of Sciences of Azerbaijan Republic [6-7].

Furthermore, the obtained formula has been proved to be valid for $q = 1$, i.e., for SO(p,l), $p \geq 2$ groups.

This work is prolongation of the publication [12]. Unfortunately, at printing paper [12] some errors were supposed. In this publication we reduce rectified expressions for some formulas [12], and also new expressions for zonal functions.

### 2. Most degenerate irreducible unitary representations of SO(p,q) groups

The most degenerate representations of groups SO(p,q), i.e., connected components of units of groups of motions of $(p + q)$-dimensional vector space holding invariant quadratic form:

$$[k,k] = k_1^2 + k_2^2 + \cdots + k_q^2 - k_{q+1}^2 - \cdots - k_{p+q}^2,$$

are given by a complex number, $\sigma$, and a number, $\varepsilon$, which take the values 0 and 1, and are constructed in the space of homogeneous functions, F(.), of given parity and defined on the cone $[k, k] = 0$, $[1 — 6]$.

Let us denote by $D^{(\sigma,\varepsilon)}$ the space of infinitely differentiable functions, $F(.)$, defined on the cone $[k,k] = 0$ without the point $k = 0$ and satisfying the following condition:

$$F(ak) = |a|^\sigma F(k) sign^\varepsilon(a), \quad a \neq 0, \quad \varepsilon = 0, 1,$$

where $a$ is any real number:

The action of the operator $T^{(\sigma,\varepsilon)}(g)$ of SO(p,q) in the space $D^{(\sigma,\varepsilon)}$ is determined as follows:

$$T^{(\sigma,\varepsilon)}(g)F(k) = F(g^{-1}k), \quad g \in SO(p,q). \tag{1}$$

The space of the representation and the representation itself may have different realizations. Let us consider one of them.

Let us introduce the spherical system of coordinates on the cone $[k, k] = 0$:

$$k = \omega(\cos\phi, \eta \sin\phi, \cos\chi, \xi \sin\chi) \tag{2}$$

where $0 < \omega < \infty$, and $\eta$ and $\xi$ are $(q-1)$-dimensional and $(p-1)$-dimensional unit vectors, respectively.

Spherical angles $\phi$ and $\chi$ change within the following limits:
$0 \leq \phi \leq \pi$, $0 \leq \chi \leq \pi$, **if** $p \geq q \geq 3$;
$0 \leq \phi < 2\pi$, $0 \leq \chi \leq \pi$, if $p \geq 2, q = 2$ (in this case $\eta$ is reduced to a constant value which we take as $\eta = 1$);
$0 \leq \phi < 2\pi$, $0 \leq \chi < 2\pi$,, if $p = q = 2$ (in this case $\eta$ and $\xi$ are reduced to a constant values which we take as $\eta = \xi = 1$).

Let us consider the reduction of functions from $D^{(\sigma,\varepsilon)}$ on the cross-section, $\omega = 1$, of the cone $[k,k] = 0$:

$$f(\phi, \eta; \chi, \xi) = F(k)\big|_{\omega=1}. \tag{3}$$

Then, in virtue of homogeneity of F(.) we obtain:

$$F(k) = \omega^\sigma f(\phi, \eta; \chi, \xi) \tag{4}$$

From equations (3)-(4) it is evident that function f(.) also has the given parity $\varepsilon$:

$$f(\phi, \eta; \chi, \xi) = (-1)^\varepsilon f(\pi - \phi, -\eta; \pi - \chi, -\xi)$$

The mappings (3)-(4) establish a one-to-one correspondence between $D^{(\sigma,\varepsilon)}$ and the space of infinitely differentiable functions defined on $S^q \otimes S^p$. From Eq. (l)-(4) we obtain the following expression, using the same notation for the space and the operator ($D^{(\sigma,\varepsilon)}$ and $T^{(\sigma,\varepsilon)}$, respectively) of the representations:

$$T^{(\sigma,\varepsilon)}(g)f(\phi, \eta; \chi, \xi) = (\omega_g / \omega)^\sigma f(\phi_g, \eta_g; \chi_g, \xi_g), \tag{5}$$

where $\omega_g, \phi_g, \eta_g, \chi_g, \xi_g$ are found from the relation of: $g^{-1}k = k_g$. In the space $D^{(\sigma,\varepsilon)}$, let us introduce the scalar product[1]:

---
[1] The bar denotes complex conjugation.

$$(f_1, f_2) = \int f_1(\phi, \eta; \chi, \xi)\overline{f_2(\phi, \eta; \chi, \xi)}(\sin\phi)^{q-2} d\phi (\sin\chi)^{p-2} d\chi (d\eta)(d\xi), \quad (6)$$

where $(d\eta)$ and $(d\xi)$ are normalized measures on $S^{q-1}$ and $S^{p-1}$, respectively, [3].

In the case q=2 or p=q=2, the differentials $(d\eta)$ or $(d\eta)(d\xi)$ are omitted because they are constants, and the other variables are integrated over the whole region.

It follows directly form Eq. (5)-(6) that the scalar product (6) is invariant for $\operatorname{Re}\sigma = -\dfrac{p+q-2}{2}$, $\varepsilon = 0,1$. By filling the space $D^{(\sigma,\varepsilon)}$ with the scalar product (6), we obtain the Hilbert space, $H^{(\sigma,\varepsilon)}$, and the most degenerated IUR's of the continuous principal series of the group $SO(p, q)$. It is easy to show that $(a, e)$ and $(2 - p - q - \sigma, \varepsilon)$ representations are unitarily equivalent [6].

Formulas (5)-(6) allow the integral representations for matrix elements of the operator of hyperbolic rotations on the surface $(k_1, k_{q+1})$ in canonical basis to be determined [3]. With respect to the canonical basis vector (which is invariant under the subgroup $S^q \otimes S^p$ the matrix elements which are the zonal spherical functions of group $SO(p,q)$ as determined in [3] are of interest. Denoting these functions by $Z_\sigma^{[p,q]}(\alpha)$ we obtain from Eq. (5)-(6) the following integral representations:

1. In the case of $p \geq q \geq 3$:

$$Z_\sigma^{[p,q]}(\alpha) = \frac{\Gamma(p/2)\Gamma(q/2)}{\pi \Gamma\left[\dfrac{p-1}{2}\right]\Gamma\left[\dfrac{q-1}{2}\right]} \int_{-1}^{+1}\int_{-1}^{+1} \theta^{\sigma/2}(1-x^2)^{\frac{p-3}{2}}(1-y^2)^{\frac{q-3}{2}} dxdy; \quad (7)$$

2. In the case of $p \geq 3$, $q = 2$:

$$Z_\sigma^{[p,2]}(\alpha) = \frac{\Gamma(p/2)}{\pi^{3/2}\Gamma\left[\dfrac{p-1}{2}\right]} \int_0^{2\pi}\int_{-1}^{+1} \Theta^{\sigma/2}(1-x^2)^{\frac{p-2}{2}} d\phi dx; \quad (8)$$

3. In the case of p=q=2:

$$Z_\sigma^{[3,2]}(\alpha) = \frac{1}{4\pi^2}\int_0^{2\pi}\int_0^{2\pi} \Theta^{\sigma/2} d\phi d\chi. \quad (9)$$

In (7)-(9) we used the following notation:

$$\Theta = (\cos\phi\, ch\alpha - \cos\chi\, sh\alpha)^2 + \sin^2\phi = 1 + (x^2 + y^2)sh^2\alpha - 2xy\, sh\alpha\, ch\alpha, \quad (10)$$

$$x = \cos\chi, \quad y = \cos\phi.$$

It is to be noted that zonal function exists only for even representations ($\varepsilon = 0$).

### 3. The canonical basis of IUR's of group SO(p,q)

The canonical basis of most degenerate IUR's of SO(p,q) groups is constructed using the results of [3]. It follows that the elements of canonical basis can be represented in the following form:

1. In the case of $p \geq q \geq 3$

$$\Xi^{(\sigma,\varepsilon)}_{\lambda l L, \mu m M}(\phi, \eta; \chi, \xi) = a^p_{\lambda l \mu m} C^{l+\frac{q-2}{2}}_{\lambda - l}(\cos\phi)\sin^l\phi \, \Xi^l_L(\eta) C^{m+\frac{p-2}{2}}_{\mu - m}(\cos\chi)\sin^m\chi \, \Xi^m_M(\xi),$$

$$\lambda \geq l \geq 0, \quad \mu \geq m \geq M; \tag{11}$$

2. In the case of $p \geq 3$, $q = 2$:

$$\Xi^{(\sigma,\varepsilon)}_{\lambda \mu m M}(\phi; \chi, \xi) = a^p_{\mu m} C^{m+\frac{p-2}{2}}_{\mu - m}(\cos\chi)\sin^m\chi \, \Xi^m_M(\xi) e^{i\lambda\phi},$$

$$\mu \geq m \geq 0; \tag{12}$$

3. In the case of $p = q = 2$:

$$\Xi^{(\sigma,\varepsilon)}_{\lambda \mu}(\phi, \chi) = \frac{1}{2\pi} e^{i(\lambda\phi + \mu\chi)}. \tag{13}$$

In formulas (14)-(15) the following notations are adopted:

$\lambda$, $\mu$, $l$, $ò$ are integers;

L, M are multi-indices which are non-negative integers;

$\Xi^l_L(\eta), \Xi^m_M(\xi)$ are the elements of canonical basis of $S(p-1)$ and $SO(q-1)$ groups, respectively, [3];

$a^{pq}_{\lambda l \mu m}, a^p_{\mu m}$ are normalization multipliers which are selected in such a way that the elements of canonical basis (11)-(13) form an orthonormalized system with respect fo the scalar product (6):

$$a^{pq}_{\lambda l \mu m} = \frac{\Gamma\left(l + \frac{q-2}{2}\right)\Gamma\left(m + \frac{p-2}{2}\right)}{2^{4-l-m-(p+q)/2}\pi} \left[\frac{(\lambda - l)!(\mu - m)!(2\lambda + q - 2)(2\mu + p - 2)}{\Gamma(\lambda + l + q - 2)\Gamma(\mu + m + p - 2)}\right]^{1/2}, \tag{14}$$

$$a^p_{\mu m} = \frac{1}{\pi}\Gamma\left(m + \frac{p-2}{2}\right)\sqrt{2^{p+2m-5}\frac{(\mu - m)!(2\mu + p - 2)}{\Gamma(\mu + m + p - 2)}}. \tag{15}$$

Furthermore, there is a restriction for the parity of the representation $\varepsilon$:

$$\lambda + \mu = \varepsilon(\mod 2). \tag{16}$$

The integral representations for matrix elements of the operator of hyperbolic rotations on the plane $(k_1, k_{q+1})$ in canonical basis, particularly for associated functions, can be written using formulas (5)-(6) and (11)-(13). Assuming the notations $P^{[pq]}_{\sigma\lambda\mu}(\alpha)$ for associated functions, we obtain the following integral representations using group-theoretical methods:

1. In the case of $p \geq q \geq 3$:

$$P_{\sigma\lambda\mu}^{[pq]}(\alpha) = a_{\lambda\mu}^{pq} \int_{-1}^{+1} \int_{-1}^{+1} \Theta^{\sigma/2} C_\mu^{\frac{p-2}{2}}(x) C_\lambda^{\frac{q-2}{2}}(y) (1-x^2)^{\frac{p-3}{2}} (1-y^2)^{\frac{q-3}{2}} dxdy; \quad (17)$$

2. In the case of $p \geq 3$, $q = 2$:

$$P_{\sigma\lambda\mu}^{[p2]}(\alpha) = a_\mu^p \int_0^{2\pi} \int_{-1}^{+1} \Theta^{\sigma/2} C_\mu^{\frac{p-2}{2}}(x) e^{-i\lambda\phi} (1-x^2)^{\frac{p-3}{2}} d\phi dx; \quad (18)$$

3. In the case of $p = q = 2$:

$$P_{\sigma\lambda\mu}^{[22]}(\alpha) = \frac{1}{4\pi^2} \int_0^{2\pi}\int_0^{2\pi} \Theta^{\sigma/2} e^{-i(\lambda\phi+\mu\chi)} d\phi d\chi; \quad (19)$$

where we have used the following designations:

$$a_{\lambda\mu}^{pq} = 4\pi^2 a_\mu^p a_\lambda^q, \quad (20)$$

$$a_\mu^p = \Gamma\left[\frac{p-2}{2}\right] \sqrt{\frac{2^{p-6} \mu!(2\mu+p-2)\Gamma(p/2)}{\Gamma(\mu+p-2)\Gamma\left[\frac{p-1}{2}\right]\pi^{7/2}}} \quad (21)$$

Remember that $\Theta$-function is determined by Eq. (10).

It is to be noted that the associated functions exist only for even $(\varepsilon = 0)$ representation as in the case of zonal functions. This follows directly from their definition. Particularly, in order to account for the condition (16), we must assume that:

$$\lambda = \nu + 2r; \quad \mu = \nu + 2s; \quad \nu = 0,1 \quad (22)$$

where $r, s$ are positive integers and introduce the following function:

$$\wp_{\sigma rs}^{pq\nu}(\alpha) = P_{\sigma\lambda\mu}^{[pq]}(\alpha) \quad (23)$$

The restrictions on the possible values of $r$ and $s$ are the same as the restrictions for $\lambda$ and $\mu$.

## 4. Associated functions and Horn's functions

The main formulas used for the calculation of zonal functions of SO(p,q) groups are the integral representations (7)-(9) and the Taylor expansions for the function $\Theta^{\sigma/2}$ given as:

$$\Theta^{\sigma/2} = \sum_{\nu=0}^{1} \frac{(-\sigma x y th\alpha)^\nu}{ch\alpha} \sum_{l=0}^{\infty} \frac{(\nu+1/2)_l}{l!} th^{2l}\alpha F_2\left(\nu - \frac{\sigma}{2}, -l, -l; \nu + \frac{1}{2}, \nu + \frac{1}{2}; x^2, y^2\right) \quad (22)$$

Here we use the following notation for Pochhammer's symbols, $(a)_n$, and Appell's functions of a second kind [8]:

$$(a)_n = \Gamma(a+n)/\Gamma(a),$$

$$F_2\left(v-\frac{\sigma}{2},-l,-l;v+\frac{1}{2},v+\frac{1}{2};x^2,y^2\right)=\sum_{m,n=0}^{l}\frac{(v-\sigma/2)_{m+n}(-l)_m(-l)_n}{(v+1/2)_m(v+1/2)_n m!n!}x^{2m}y^{2n}.$$

Series (22) converges uniformly and absolutely for sufficiently small values of α, namely at $ch2\alpha < 3$ (the sufficient condition !).

In order to find the associated functions of SO(p, q) groups it is necessary to substitute the Taylor expansion for the function $\Theta^{\sigma/2}$ into Eq.(17)-(19) and to perform the integration. As a result we obtain:

$$\wp_{\sigma rs}^{pqv}(\alpha)=\frac{(-1)^{s+r+v}}{ch\alpha}A_1 A_2 \sum_{l=\max(s,r)}^{\infty}\frac{l!(v+1/2)_l}{(l-s)!(l-r)!}F_2(s+r+v-\sigma/2, s-l, r-l;$$

$$2s+v+p/2, 2r+v+q/2;1,1)(th\alpha)^{2l+v}, \qquad (23)$$

where

$$A_1=\frac{2^{v-3(p+q)/2}(-\sigma/2)_{s+r+v}}{\Gamma(2s+v+p/2)\Gamma(2r+v+q/2)}\left[\frac{\pi\Gamma(2s+v+p-1)\Gamma(2r+v+q-1)\Gamma(p/2)\Gamma(q/2)}{(2s+v)!(2r+v)!\Gamma\left[\frac{p-1}{2}\right]\Gamma\left[\frac{q-1}{2}\right]}\right]^{1/2}$$

$$A_2=\sqrt{\frac{\left(2s+v\frac{p-2}{2}\right)\left(2r+v+\frac{q-2}{2}\right)}{(2s+v+p-2)(2r+v+q-2)}}.$$

Eq. (23) for associated functions of SO(p,q) groups can be rewritten more compactly using generalized hypergeometric functions of the two variables, (A1)-(A2):

$$\wp_{\sigma rs}^{pqv}(\alpha)=A_1 A_2 A_3 \frac{(th\alpha)^{2s+v}}{ch\alpha}{}_5F_3\begin{bmatrix}11 & s+1, & s+v+1/2 \\ 10 & s+r+v-\omega/2 & \\ 01 & \dfrac{q+\sigma}{2}, & s-r+\dfrac{p+\sigma}{2} \\ 11 & 2s+v+p/2, & s+r+q/2 \\ 01 & 1+s-r & \end{bmatrix}th^2\alpha, th^2\alpha$$

$$A_3=\frac{(2s+v)!\left[\dfrac{2-\sigma-q}{2}\right]_{s-r}}{(s-r)!4^s}. \qquad (24)$$

Formula (28) is valid for $s \geq r$. For $s \leq r$ the expression for associated function is derived from (24) using the rearrangement:

$$\begin{bmatrix}p & q & r & s \\ q & p & s & r\end{bmatrix}.$$

From this, one can attain the symmetry property of the associated function for the groups SO(p, q):

$$\wp_{\sigma rs}^{pqv}(\alpha)=\wp_{\sigma sr}^{qpv}(\alpha) \qquad (25)$$

It is also important to note that formulas (23)-(25) are true for all

$SO(p,q)$, $p \geq 2, q \geq 2$ groups.

Assuming $s = r = v = 0$ **in** (23)-(25), the expressions for zonal functions of $SO(p,q)$, $p \geq 2, q \geq 1$ groups can be derived as:

$$Z_\sigma^{[pq]}(\alpha) = \frac{1}{ch\alpha} \sum_{l=0}^{\infty} \frac{(1/2)_l}{l!} F_2(\sigma/2, -l, -l; p/2, q/2; 1, 1)(th\alpha)^{2l}, \qquad (26)$$

$$Z_\sigma^{[pq]}(\alpha) = \frac{1}{ch\alpha} {}_5F_3 \begin{bmatrix} 11 & 1/2, & 1 \\ 01 & -\sigma/2 & \\ 10 & \dfrac{q+\sigma}{2}, & \dfrac{p+\sigma}{2} \\ 11 & p/2, & q/2 \\ 01 & 1 & \end{bmatrix} th^2\alpha, th^2\alpha \Bigg]. \qquad (27)$$

The advantages of Eq. (26)-(27) involve the fact that symmetry between $p$ and $q$ evident.

The following expansions are obtained from Eq. (17)-(19) for the function $\Theta^{\sigma/2}$:

$$\Theta^{\sigma/2} = \frac{\pi \Gamma\left[\dfrac{p-1}{2}\right]\Gamma\left[\dfrac{q-1}{2}\right]}{\Gamma(p/2)\Gamma(q/2)} \sum a_{\lambda\mu}^{pq} P_{\sigma\lambda\mu}^{[pq]}(\alpha) C_\mu^{\frac{p-2}{2}}(x) C_\lambda^{\frac{q-2}{2}}(y), \qquad (28)$$

$$\Theta^{\sigma/2} = \frac{2\pi^{3/2} \Gamma\left[\dfrac{p-1}{2}\right]}{\Gamma(p/2)} \sum a_\mu^{pq} P_{\sigma\lambda\mu}^{[p2]}(\alpha) C_\mu^{\frac{p-2}{2}}(x) e^{i\lambda\phi}, \qquad (29)$$

$$\Theta^{\sigma/2} = \sum P_{\sigma\lambda\mu}^{[22]}(\alpha)(x) e^{i(\lambda\phi+\mu\chi)}, \qquad (30)$$

where summations are performed over all admissible values, À, /ё, which satisfy the restriction $\lambda + \mu = 0 (\mathrm{mod}\, 2)$.

Expansions (28)-(30) allow supplementary functional relationships to be obtained for the spherical functions for SO(p,q) groups, and these may be useful in the solutions of many problems.

It should be noted that the convergence of the series in Eq. (28)-(30) is realized in the sense of convergence of distributions over Frechet spaces [9]:

$$C^\infty([-1,1]\times[-1,1]), C^\infty([-1,1]\times[0,2\pi]) \text{ and } C^\infty([0,2\pi]\times[0,2\pi]),$$

respectively[2].

To conclude, it is to be noted that we have studied the functions of $SO(p,q)$ groups for the representations of the continuous principal series. We will proceed to study spherical functions for the representations of discrete principal and supplementary series in the future.

## 6. Appendix

The generalized hypergeometric Horn's series with $r$ variables is defined as

---

[2] In this case the ends of interval [0, 2π] must be identified.

follows [10]:

$$\sum_{n_1 \cdots n_r = 0}^{\infty} \frac{\prod_{\alpha=1}^{p} \left(a_\alpha, \sum_{j=1}^{r} u_{\alpha j} n_j\right)}{\prod_{\beta=1}^{q} \left(b_\beta, \sum_{j=1}^{r} v_{\beta j} n_j\right)} \cdot \prod_{i=1}^{r} \frac{X_i^{n_i}}{n_i!}, \qquad (A1)$$

$$\sum_{\alpha}^{p} u_{\alpha j} = \sum_{\beta}^{q} v_{\beta j} + 1, \quad 1 \leq j \leq r,$$

where

$$(\lambda, n) = \frac{\Gamma(\lambda + n)}{\Gamma(\lambda)} = (\lambda)_n, \quad (\lambda, 0) = 1.$$

It is convenient to denote the series (A1) by the notations [11]:

$$_pF_q \begin{bmatrix} u_{\alpha_1 1}, & \cdots & u_{\alpha_1 r} & \{a_{\alpha_1}\} \\ \cdots & \cdots & \cdots & \cdot \\ \cdots & \cdots & \cdots & \cdot \\ \cdots & \cdots & \cdots & \cdot \\ u_{\alpha_s 1}, & \cdots & u_{\alpha_s r} & \{a_{\alpha_s}\} \\ v_{\beta_1 1}, & \cdots & v_{\beta_1 r} & \{b_{\beta_1}\} \\ \cdots & \cdots & \cdots & \cdot \\ \cdots & \cdots & \cdots & \cdot \\ \cdots & \cdots & \cdots & \cdot \\ v_{\beta_t 1}, & \cdots & v_{\beta_t r} & \{b_{\beta_t}\} \end{bmatrix} X_1, \cdots X_r , \qquad (A2)$$

$$s \leq p, \quad r \leq q,$$

where the multiples in the numerator (denominator) of (A2), corresponding to same values of $u_{\alpha j}$ $(v_{\beta j})$, $1 \leq j \leq r$, are unified in the same row of (A2), i.e., the set of the same values of $\alpha, 1 \leq \alpha \leq p$ $(\beta, 1 \leq \beta \leq q)$ are decomposed to the subsets $\alpha_1, \ldots, \alpha_s$ $(\beta_1, \ldots, \beta_t)$ with the same values $u_{\alpha j}$ $(v_{\beta j})$, $1 \leq j \leq r$.